%% file: RealTimeAnalysisLHC.tex
\documentclass[twoside,11pt]{article}

\usepackage{jmlrwcp2e}
\usepackage{url}
\usepackage{ifthen}
\newboolean{pdflatex}
\setboolean{pdflatex}{true}

\newboolean{articletitles}
\setboolean{articletitles}{true} 

\newboolean{uprightparticles}
\setboolean{uprightparticles}{false} 

\newboolean{inbibliography}
\setboolean{inbibliography}{false} 
\usepackage{rotating}
\usepackage{multirow}
\usepackage{geometry}
\usepackage{afterpage} 
\usepackage{microtype}
\usepackage{lineno}  
\usepackage{xspace} 
\usepackage{caption} 
\usepackage{graphicx}  
\usepackage{color}
\usepackage{colortbl}
\usepackage{amsmath} 
\usepackage{amssymb}
\usepackage{amsfonts}
\usepackage{upgreek} 
\usepackage{hyperref}

\input{lhcb-symbols-def}

\jmlrheading{42}{2015}{1-18}{}{}{Vladimir V. Gligorov}{NIPS 2014, HEP-ML Workshop}

\ShortHeadings{Real-time data analysis at the LHC}{Gligorov}
\firstpageno{1}

\begin{document}

\title{Real-time data analysis at the LHC: present and future}

\author{\name Vladimir V. Gligorov \email vladimir.gligorov@cern.ch \\
       \addr CERN\\
       CH-1211\\
       Switzerland}

\editor{Glen Cowan, C\'ecile Germain, Isabelle Guyon$^\text{d}$,
Bal\'azs K\'egl,and David Rousseau}

\maketitle

\begin{abstract}
The Large Hadron Collider (LHC), which collides protons at an energy of 14 TeV, produces hundreds of exabytes of data per year,
making it one of the largest sources of data in the world today. At present it is not possible to even transfer most of this
data from the four main particle detectors at the LHC to ``offline'' data facilities, much less to permanently store it for future processing.
For this reason the LHC detectors are equipped with real-time analysis systems, called triggers, which process this volume of
data and select the most interesting proton-proton ($pp$) collisions. The LHC experiment triggers reduce the data produced by
the LHC by between 1/1000 and 1/100000, to tens of petabytes per year, allowing its economical storage and further analysis.
The bulk of the data-reduction is performed by custom electronics which ignores most of the data in its decision making,
and is therefore unable to exploit the most powerful known data analysis strategies. I cover the present
status of real-time data analysis at the LHC, before explaining why the future upgrades of the LHC experiments will increase
the volume of data which can be sent off the detector and into off-the-shelf data processing facilities (such as CPU or GPU farms)
to tens of exabytes per year. This development will simultaneously enable a vast expansion of the physics programme of the
LHC's detectors, and make it mandatory to develop and implement a new generation of real-time multivariate analysis tools in
order to fully exploit this new potential of the LHC. I explain what work is ongoing in this direction and
motivate why more effort is needed in the coming years.
\end{abstract}


\section{Introduction}
\label{sec:introduction}

When hearing the phrase ``real-time data analysis'', three questions immediately spring to mind. Firstly, what does
``real-time'' mean? Secondly, what does ``data'' mean? And thirdly, what does ``analysis'' mean? All of these terms can,
and do, mean quite different things in different scientific domains. For this reason, it is important to begin these proceedings
with some definitions and context, to make the discussion which follows more understandable to readers who are not high energy physicists.
I apologize to readers who \textbf{are} high energy physicists if the style of this paper consequently seems overly pedagogical.

I will be discussing real-time data analysis in the context of the Large Hadron Collider (LHC), a particle accelerator which
nominally\footnote{During the 2010-2012 period, the LHC ran at a reduced centre-of-mass energy of first 7, then 8~TeV, and a reduced collision frequency
of 15~MHz.} collides beams of protons at a centre-of-mass energy of 14~TeV and a rate of 40 million per second (MHz). For readers who are not high
energy physicists, two factoids can help to set the scale of the problem: each beam of protons carries the energy of a fully loaded
TGV train travelling at $\approx 150$~kmh$^{-1}$, and the radiation pressure at the centre of the collision is similar to resting the sun on the head
of a pin. There are four main detectors of particles at the LHC: ALICE, ATLAS, CMS, and LHCb, whose goal is to record the trajectories
of the particles produced in these $pp$ collisions, identify them, measure their energies, and consequently make inferences about their underlying physical properties. 
These collaborations bring together close to 10,000 physicists in pursuit of this goal.

The essential difficulty faced by the LHC experiments is the volume of the data produced. For example, each $pp$
collision in the LHCb experiment results in around 50~kB of data, or a nominal data rate of around 1.5~TB per second. For ATLAS and CMS these
data rates are an order of magnitude greater. As the LHC (\cite{Pojer:2012zz}) takes data for around $4\cdot 10^6$~seconds per year, the detectors process on
the order of 100~EB of data each year, on a par with the largest commercial data processing applications. Two problems arise from these facts.
Firstly, economically transferring Terabytes of data per second is only just becoming possible, while the detectors were designed over a decade ago, when it was not.
And secondly, even if the data could be transferred off the detectors, it is not economically feasible to store and distribute
that much data to the thousands of physicists around the world waiting to analyse it. For both of these reasons, the LHC experiments use real-time
data analysis, known as ``triggering'', in order to select roughly between $0.001\%$ and $0.01\%$ of the most interesting data collected 
by the detectors for further analysis and \textbf{permanently discard} the rest. Real-time therefore means that the analysis should be performed
rapidly enough that the available computing resources do not saturate; in other words, every $pp$ collision should be analysed and kept
or rejected on its merits, not discarded because the real-time analysis facilities were at capacity. Real-time in this context does not generally
mean a commitment to analyse the data within any given time period: as we shall see in due course, by correctly structuring the problem it is possible
to stretch the concept of ``real-time'' quite a long way.

Having defined the basic terminology, the remainder of these proceedings shall aim to give the reader an understanding of the current
state-of-the-art in real-time data processing at the LHC, as well as an idea of where this field will go in the future. The LHCb experiment, on
which I work, will be the main reference point for all examples, but I will also touch on the strategies used by the other experiments in order
to illustrate specific points. I will begin by further clarifying what is data and what is analysis to an LHC physicist;
readers who are high energy physicists may wish to skip this section. I will then describe the
current technical constraints in detail and explain how they are overcome within the context of LHCb. The limitations of the current solutions
will also be discussed, and planned improvements will be described within this context. Natural units in which $c=\hbar=1$ are used throughout.

\section{Data and analysis in high energy physics}

\begin{figure}
\centering
\includegraphics[width=0.9\linewidth]{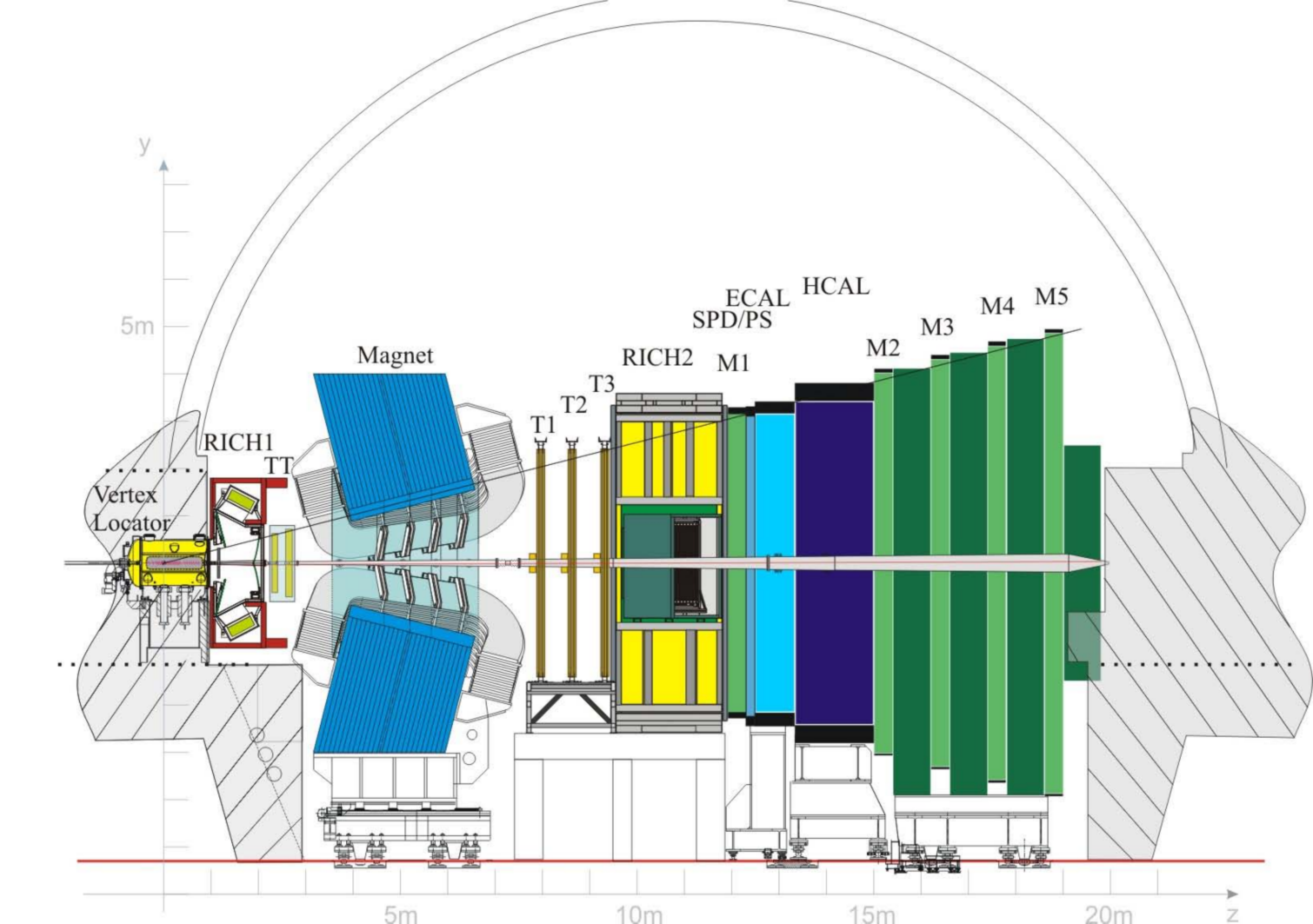}
\caption{The LHCb detector layout. The protons travel along the beamline, which is the horizontal or \textbf{z} axis on the figure,
and collide in the ``Vertex Locator'' on the left.}
\label{fig:gene-2008-002_01}
\end{figure}

In order to get a better idea of what data and analysis mean to high energy physicists,
it will be useful to examine a specific LHC detector, namely LHCb.
The \lhcb detector (\cite{Aaij:2014jba}), shown in Fig.~\ref{fig:gene-2008-002_01} is a single-arm forward
spectrometer designed for studing beauty and charm (heavy flavour) hadrons produced in $pp$ collisions
at the \lhc at CERN. In what follows ``transverse'' means transverse to the LHC beamline,
which is the horizontal or \textbf{z} axis on the figure, ``downstream'' means to the right of the figure and
``upstream'' to the left.

The LHCb detector consists of several sub-detectors, whose names are labelled
on the figure. The protons travel along the beamline and collide in the ``Vertex Locator'' (VELO) on the left,
which reconstructs 
the positions of the $pp$ collisions and the trajectories of the charged
particles produced in these collisions. The reconstruction of charged particle trajectories is commonly
referred to as ``tracking''. As there is no magnetic field inside the VELO all tracks are straight lines,
but to measure their momentum the charged particles must pass through the magnet. The bending
of their trajectories is proportional to the track momentum, while the charge of the track determines the
direction in which it bends. LHCb's dipole
magnet, with a bending power of about $4{\rm\,Tm}$, is therefore surrounded by one upstream (TT) and three downstream (T1~-~T3)
tracking stations. The combined tracking system has a momentum resolution
$\Delta p/p$ that varies from 0.4\% at 5\gevc to 0.6\% at 100\gevc,
and an impact parameter\footnote{Impact parameter is the transverse distance of closest
approach between a track and a vertex, most commonly the primary $pp$ interaction vertex.}
resolution of 20\mum for tracks with high
transverse momentum. The decay times of particles which live long enough to decay away from the primary $pp$
interaction vertex can be reconstructed with a resolution of 50~fs. 

In addition to determining the momenta of charged particles, it is necessary to determine
their mass, which is the single clearest discriminant between different types of particles
produced at the LHC. This is done using two 
ring-imaging Cherenkov detectors, labeled RICH1 and RICH2 on Fig.~\ref{fig:gene-2008-002_01}.
These use the Cherenkov effect, in which particles travelling faster than the local speed of light
in a material (in LHCb's case, a gas) emit a ``shock-cone'' of light whose opening angle is proportional to the particle's speed,
to measure the speed of charged particles. Combined with the existing momentum measurement, this determines
the charged particle masses.

Neutral particles, which do not bend in a magnetic field, are identified, and their energy measured, by a calorimeter system consisting of
scintillating-pad and pre-shower detectors (SPD/PS), an electromagnetic
calorimeter (ECAL), and a hadronic calorimeter (HCAL). Finally, the only particles which penetrate
beyond the HCAL are muons, which are identified by a muon
system (M1--M5) composed of alternating layers of iron and multiwire
proportional chambers. A summary of the basic particles and their properties is given in Tab.~\ref{tab:parts} for convenience.

\begin{table}[htbp]
\centering
\caption{Basic particles reconstructed in the LHCb detector and their properties.}
\vspace{0.25cm}
\label{tab:parts}
  \begin{tabular}{l|l}  
      \hline 
    Particle & Summary \\    
    \hline 
    \multirow{2}{*}{Electron} & Electrically charged, detected as a trajectory in the \\
                              & tracking system ending in the electromagnetic calorimeter.\\
    \hline
    \multirow{3}{*}{Muon}     & Electrically charged and able to penetrate through the \\ 
                              & calorimeter system. Detected as a trajectory in the tracking \\
                              & system extending into the muon chambers. \\
    \hline
    \multirow{5}{*}{Pion, Kaon, Proton}     & Electrically charged, detected as trajectories in the tracking\\
                                            & system ending in the hadronic calorimeter. These three particle\\
                                            & types are distinguished by associating photon rings in the RICH \\
                                            & detectors to their trajectories and measuring the opening angle\\
                                            & of this light cone.\\
    \hline
    \multirow{3}{*}{Photon}   & Electrically neutral, identified by an energy deposit\\
                              & in the electromagnetic calorimeter with no nearby\\ 
                              & charged particle trajectory. \\
    \hline 
  \end{tabular}
\end{table}

While the layout of the LHCb detector, facing only one direction away from the $pp$ collision point,
is unusual, the described sub-detectors are quite typical for a high-energy physics experiment. The data collected
by the detector consists of electrical signals generated when the particles
produced in the $pp$ collision pass through the sub-detectors. These electrical signals are converted by the detector's
read-out electronics into ``hits'' and/or ``clusters'' (of hits), corresponding to the point at which a certain particle
traversed a certain region of a sub-detector. This lowest-level data is referred to as the ``raw event''.
Subsequently there are many levels of ``analysis'' which can be applied to the raw event:
for example the aforementioned reconstruction of charged particle trajectories, the reconstruction of the $pp$ interactions,
or the combination of many charged and/or neutral particles into other particles. When high-energy physicists talk about
data analysis they typically mean the measurement of some parameter of interest, for example a fundamental constant of nature such as the mass or lifetime of
a given particle. In this context, analysis generally consists of making particle candidates, using their properties to distinguish
between signal and backgrounds, and then carrying out an appropriate statistical treatment (e.g. a regression using appropriate variables) in
order to extract a value and uncertainty for the parameter of interest. This description, however, hides the steps which are required
to go from the raw event to the high-level quantities used in an analysis, steps which are often called ``reconstruction'' by physicists,
but in fact have a myriad of complex and often ambiguous names within the field. In order to avoid this kind of jargon as much as possible, 
I will simply refer to the general process of extracting parameters of interest from the raw event as analysis, while
reconstruction and selection will be used, as in this section, in their natural English language sense. Any given analysis will contain some
number of reconstruction steps\footnote{Including zero---high energy physicists do not generally attempt to extract parameters of interest directly
from the raw event, but this is a question of what works best, not a logical impossibility.}, followed by a statistical treatment to extract
the parameter of interest. 

\section{Why is real-time analysis necessary at the LHC?}

We are now in a position to answer the question of why LHC physicists
have to perform at least part of their data analysis in real-time. As explained in the introduction,
the LHC collides protons at an event rate of 40~MHz, for around $4\cdot 10^6$~seconds per year. In fact each such event
consists of a crossing of the two proton beams, during which one or more $pp$ interactions can occur.
The number of $pp$ interactions per event\footnote{This number is typically referred to as ``pileup'' by high energy physicists.}
can be tuned according to what is optimal for any given detector: more
interactions between protons means that more interesting particles can be produced, but also that the complexity
of the raw event increases and, consequently, that the efficiency of any reconstruction or selection applied to this
raw event decreases. In the case of the present LHCb detector the optimum is found to be around 1 $pp$ interaction
per event, whereas it is closer to 20 for ATLAS and CMS\footnote{These very different optima are a consequence of the
different physics aims and detector designs. In particular because LHCb looks along the beamline, where the particle density
is greatest, it would need a much more expensive detector to be able to efficiently measure events containing multiple
$pp$ interactions. This is part of the motivation behind the LHCb upgrade which will be built in 2020 and allow running
at an average of 5 $pp$ interactions per event.}.
There are now two elements which drive the need for, and nature of, real-time
analysis at the LHC: the technical constraints on the transfer of the raw data from the detector electronics to offline
computing facilities, and the cost of storing and distributing the data to physicists for further analysis.

Each LHCb raw event is around 50~kB in size, which means that the detector records around 1.5~TB of data
per second; ATLAS and CMS record around fourty times more than LHCb, while ALICE records around one order of magnitude
less\footnote{In the case of ALICE, each individual event is very large, but the collision rate is much smaller than for LHCb, ATLAS, or CMS.}.
Although such data rates may not seem impossible today, it is worth bearing in mind that these detectors were designed
at the turn of the century with the idea that they would start taking data in 2008, at which point the technology to transfer
such data rates simply wasn't available. All collaborations therefore require a part of the real-time data analysis to be implemented
in custom electronics which operate at the detector readout. In the case of LHCb this ``hardware trigger'' reduces the data rate
by a factor of $1/30$, while for ATLAS and CMS the reduction is around $1/300$, roughly in proportion to their ten times
larger input data rate. The details of what data is kept and what is discarded are given later in Sec.~\ref{sec:present}.
The real-time analysis stages following this hardware trigger are known as ``software'' or ``high-level'' triggers. 

The second constraint, on the cost of storing and distributing the data, is most easily illustrated with reference
to the LHCb detector. The LHCb detector cost (\cite{Antunes-Nobrega:630827}) around 75 million CHF to construct, while
the data storage and processing currently costs around 7 million CHF per year\footnote{See slides by Ian Bird at \href{https://indico.cern.ch/event/361440/contribution/2/material/slides/1.pdf}{https://indico.cern.ch/event/361440/contribution/2/material/slides/1.pdf}.}. By constrast, LHCb's current real-time data analysis
facility cost around 5 million CHF in total,
and has a usable lifetime of around five years after which the computing nodes begin to fail and need to be replaced. 
This facility reduces the number of $pp$ collisions read from the detector by a further factor 100, for a total
data reduction of 1 in 3000.
One can therefore crudely estimate that trying to store and distribute all the data collected by LHCb would cost around 21,000 times more
than performing a real-time analysis first and only storing the subset of data which passes this analysis\footnote{This
calculation is a little unfair because raw events which have been selected by a real-time analysis are somewhat more complex than
average and consequently take longer to process. Even assigning an unrealistic factor of 10 for such an effect, however, would not
change the basic conclusion.}. Irrespective of
whether it would be desirable or not, such an approach is clearly not feasible economically. Similar considerations apply to ATLAS, CMS,
or ALICE.

\section{Making time less real: budgeting analyses}

The technical goal behind real-time analysis at the LHC is therefore to reduce the volume of data which has to be stored and distributed.
Part of the data reduction is achieved by the aforementioned hardware triggers which reduce the data rate to what can be transferred out
of the detector electronics; as these are implemented in custom electronics and must operate at a fixed latency they do not allow
for much flexibility in how the data volume is reduced. I will describe the state-of-the art of hardware triggers in Sec.~\ref{sec:present}, but
will concentrate in this section on the more complex question of what happens to the data once it has been transferred from the detector
electronics to a processing facility. In this I am not assuming that this processing facility consists of CPU servers, although this is usually
the case. The data processing can be accomplished by whatever technology suits the problem best: CPUs, GPUs, FPGAs, in-between technologies
like the XeonPhi, etc. The crucial difference between a hardware trigger and a data processing facility is that the former must decide
to keep or reject a raw event within a fixed latency\footnote{To say this another way: compared to installing a hard disk into a CPU server,
it is not economically or technically feasible to implement
sizeable buffers within the readout electronics of detectors which would allow data to be stored while the hardware trigger makes a decision about it.}
after which the data is either read from the detector electronics or lost,
whereas the latter can be structured to take an arbitrarily long time to make a decision, as we shall now see.

\begin{figure}
\centering
\includegraphics[width=1.0\linewidth]{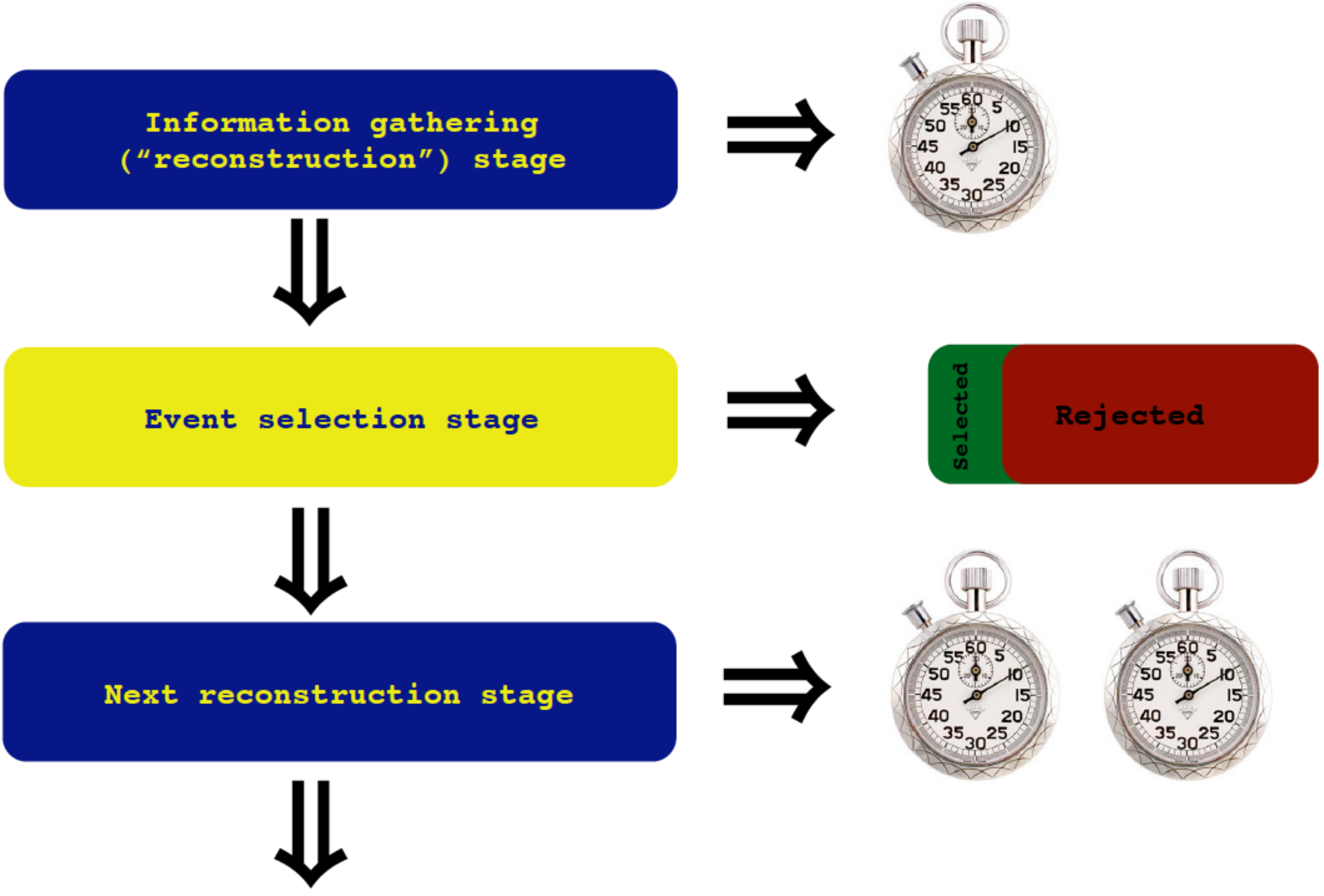}
\caption{A real-time analysis cascade, in which early analysis stages use information which can be obtained quickly (one stopwatch) in order
to reduce the data rate and enable more sophisticated and costly analyses (two stopwatches) which are also consequently more efficient. The label ``reconstruction''
here corresponds to one analysis stage, and the proportion of events kept or rejected is purely illustrative. While the cascade
is labeled in terms of event reconstruction or rejection, the logic would be the same if parts of an individual event were being kept or discarded.}
\label{fig:RealTimeSelection}
\end{figure}

Data reduction can be achieved either by selecting a subset of the raw events which has been judged
to be especially interesting, or by selecting a subset of the data within a given raw event which has been judged
to be especially interesting, or by a mixture of these two. Traditionally most high energy physics experiments have used 
real-time analysis to select a subset of interesting events, and this is\footnote{See \cite{Aaij:2012me,Aad:2012xs,Brooke:2013hnf}.}
currently the dominant data reduction strategy of LHCb, ATLAS, and CMS.
More recently\footnote{See \cite{ALICEO2,CERN-LHCC-2014-016,Fitzpatrick:1670985}.}
the idea of using real-time analysis in order to select interesting parts of a given event has been gaining traction,
pioneered by ALICE and to an extent LHCb. In either case, the data reduction always proceeds through the kind of cascade shown in Fig.~\ref{fig:RealTimeSelection}.
The goal of such a cascade is to reduce the data volume to the target level, while retaining interesting information with maximal efficiency.

The typical real-time analysis cascade used in high-energy physics interleaves reconstruction stages, which obtain information about the
event (feature-building in the language of data science),
with selection stages which decide whether to keep or discard a specific (part of an) event based on the information obtained in the preceeding
reconstruction stages. The reason for this cascade structure is that the computing budget available for real-time analysis is too small to
allow all the relevant reconstruction to be performed upfront. At the same time, the reconstruction which we \textbf{can} afford to perform
upfront does not allow for an efficient data reduction to the final target, but it \textbf{does allow} for an efficient partial data reduction.
The cascade leverages this limited ability of the fast analysis to perform a partial data reduction and consequently make more time
available for more complex analysis steps. To give a simple example, we might have an overall budget of 10~ms in which to reduce the data
by a factor 100, and two analysis steps: a fast analysis which takes 5~ms and provides information which can efficiently reduce the data by a factor 10, 
and a slow analysis which takes 50~ms and provides information which can reduce the data by a factor 100. Running the fast analysis reduces the overall
time budget by 5~ms, but also reduces the data volume by a factor 10. It thus leaves an effective budget of 50~ms for processing the data which
survives the first analysis step, enough to run the slow analysis and perform the final data reduction.

While the described cascade allows us to stretch the concept of real-time to a certain extent, it remains constrained by the logic
that the real-time analysis should run while the collider is colliding particles. However, a typical collider will only be running
around $20\%$ of the time. This fairly universal ceiling is driven by the maintenance and commissioning needs of the machine,
and opens the possibility to stretch real-time even further by temporarily buffering the data while the collider is running and
completing the processing and data reduction during the collider downtime. Such a scheme has been implemented by LHCb (\cite{LHCbDeferredTrigger})
and proposed by ALICE (\cite{ALICEO2}), and it allows for the computing facilities to be leveraged by further factors; the precise
gain depends on the specific detector and data reduction problem. 

In order to understand how such a data buffer can help to make real-time less real, let us now revisit our earlier example.
Since the data buffer's performance depends on the volume of data, we will have to add these parameters to the model: a 50~GBs$^{-1}$
input data rate, and an overall data buffer of 5~PB~\footnote{This number corresponds to the data buffer which LHCb will use during 2015--2019 datataking.}.
Our first instinct might be to attempt to run the slow analysis from the beginning,
which would mean that $80\%$ of the input data rate, or 40~GBs$^{-1}$ would need to be buffered. The 5~PB buffer would consequently allow
for around 35~hours of buffering before it filled up, while it could be emptied at a rate of 10~GB$^{-1}$. It might seem
that this approach works, since the collider is only running at $30\%$ of the time, but this neglects the fact that the collider runtime and
downtime are not uniformly distributed throughout the year: most of the LHC downtime occurs during a several-month long winter shutdown
and two-week-long ``technical stops'' distributed throughout the year. The runtime is consequently also concentrated, with a peak structure
of repeated 15~hour long collision periods with breaks of 2--3~hours. In this context our naive approach is clearly suboptimal.

\begin{figure}
\centering
\vspace{-1cm}
\includegraphics[width=1.0\linewidth]{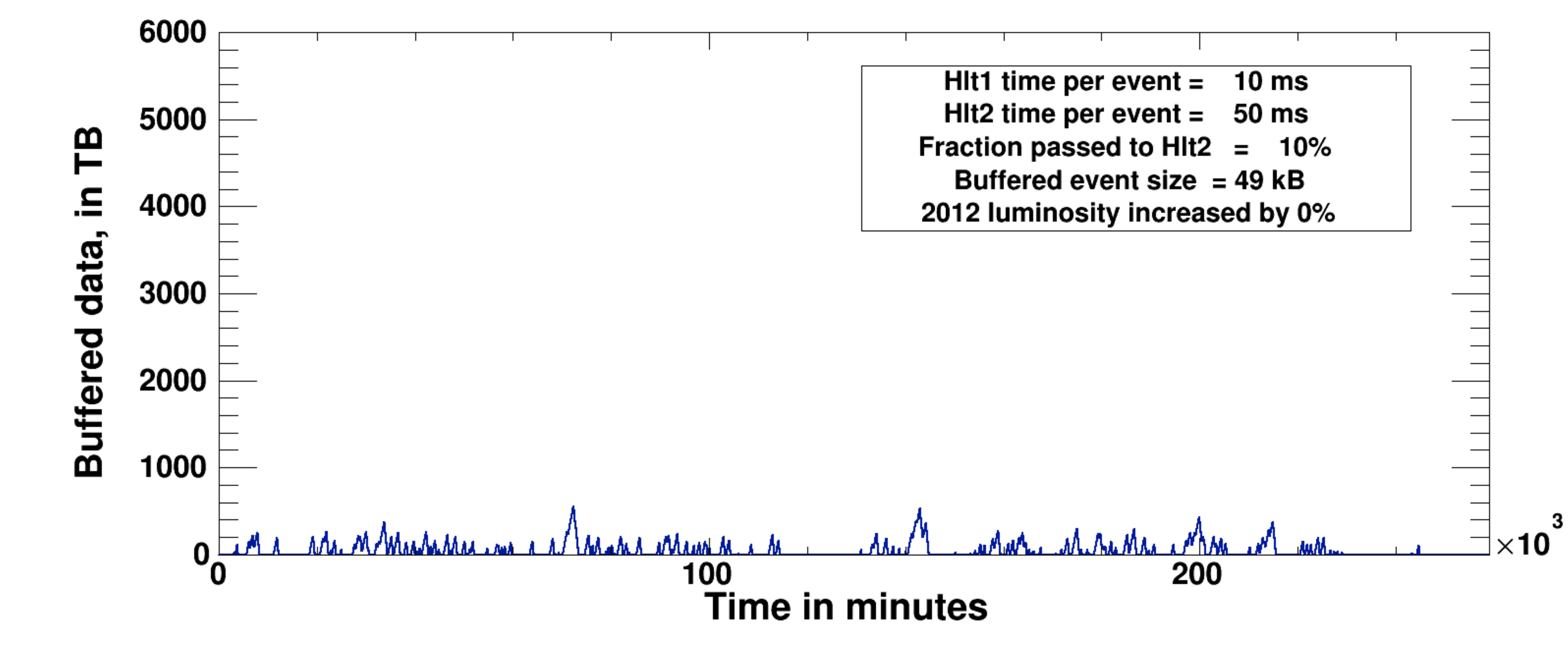}
\includegraphics[width=1.0\linewidth]{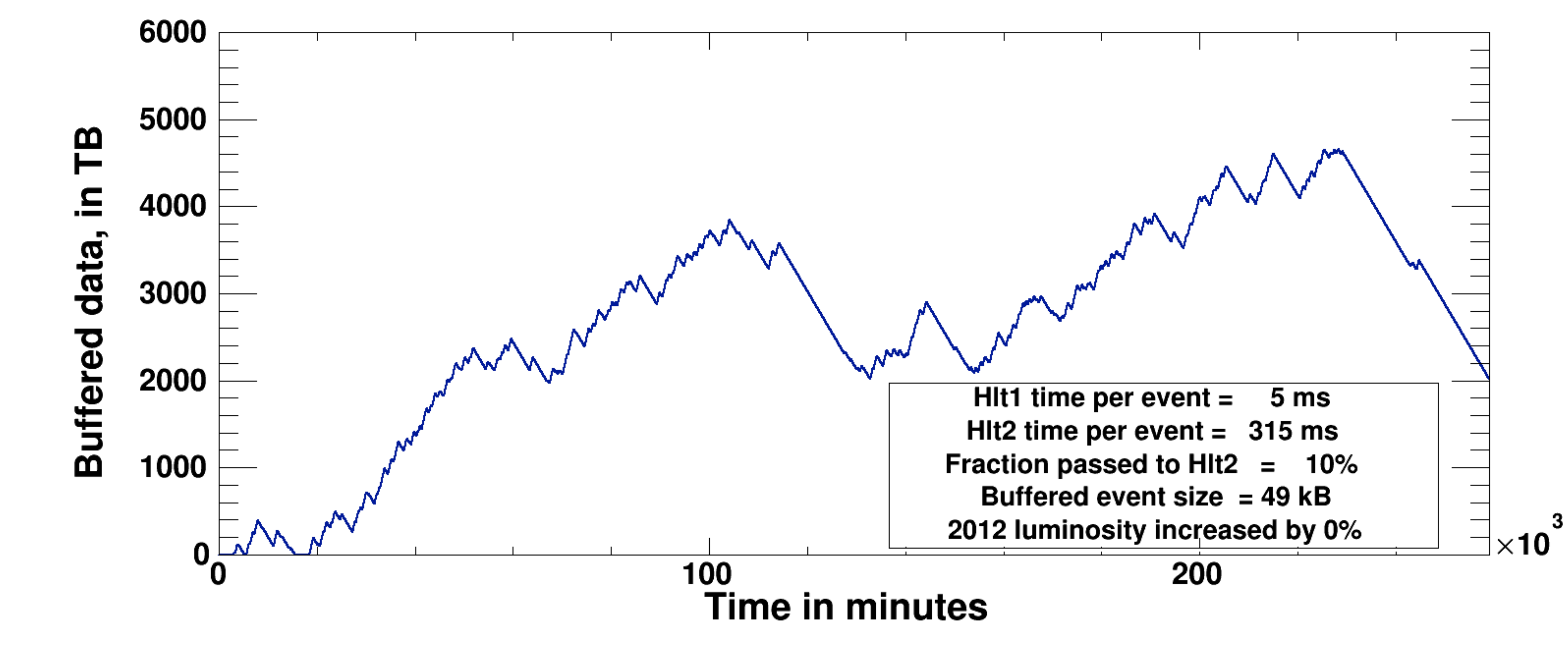}
\includegraphics[width=1.0\linewidth]{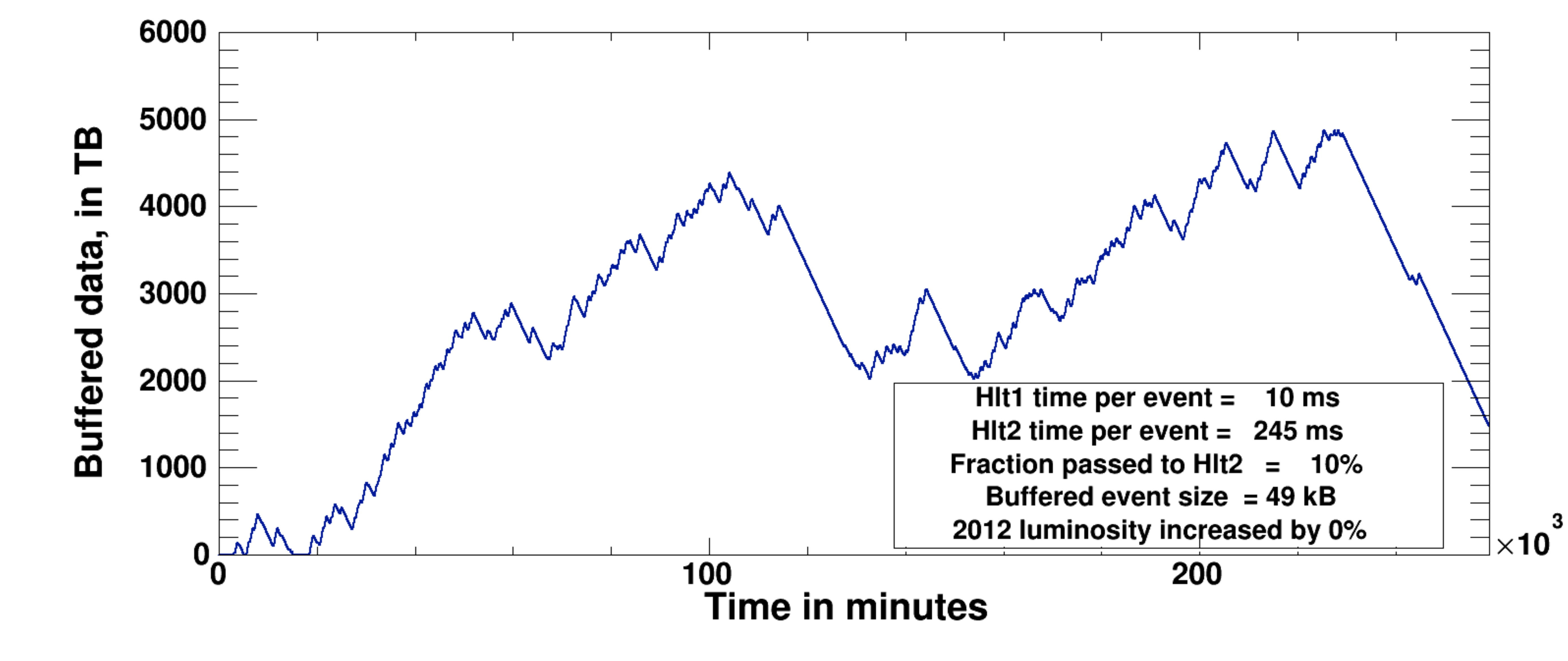}
\caption{Three possible ways to use the disk buffer: give more time to the fast analysis (top), give more time to the slow analysis (middle),
or share additional time between the two (bottom).}
\label{fig:diskbuffer}
\end{figure}

The problem with the naive approach is that it requires us to buffer too much data, so let us now try to combine our earlier analysis cascade
with the buffer by buffering only that data which survives the first analysis stage. We can then use the buffer to allow either the fast analysis,
slow analysis, or both, to take longer. In order to simulate the possible gains
we use the observed LHC fill structure in 2012, and show the buffer usage in three possible scenarios in Fig.~\ref{fig:diskbuffer}. 
When giving all the additional time to the fast analysis, the buffer is hardly used, while the amount of additional time which can be given 
to the slow analysis varies between a factor of 5 and 6, depending on whether the fast analysis is also given more time. The precise optimum
will of course be problem dependent, but it should be clear that this buffering approach allows for non-linear gains in time for the final 
analysis stages. The buffer also allows for a time delay between the fast and slow analyses, which can be used to, for example, better calibrate
the detector in order to make the later analysis stages more precise. 

It is interesting to ask how such a buffer should be optimally implemented. LHCb has chosen a configuration in which the real-time analysis
is performed on a ``farm'' of CPU servers, with the buffering implemented locally on each server. This saves on network resources compared to having
some centralized disk server to which all the processing servers send their buffered events. 
Of course one can also have multiple such buffers, as ALICE (\cite{ALICEO2}) are planning, and at some point it can become more economically efficient
to buffer some of the data in the cloud\footnote{Also known as the GRID within high energy physics, see \href{http://wlcg.web.cern.ch}{http://wlcg.web.cern.ch} for more details. It is worth
noting that in this model, the data transfer costs become comparable to the data storage costs and must be carefully optimized.}.
We will come back to the future of real-time analysis cascades and buffers in Sec.~\ref{sec:future}.

Finally, the described analysis cascades are of course not an invention of physicists. Apart from the specific problems which they are used for,
it is worth remarking that analysis cascades in high energy physics are usually constructed ``by hand'', with reconstruction and selection
steps selected by physicists and optimized by trying a few reasonable configurations. Recent work (\cite{DjalelMDDAG}) in data science
opens the possibility of an automatic, machine learning based, design of such analysis cascades. In particular, this paper showed that the MDDAG
algorithm could learn the optimal analysis cascade if given a set of discriminating features and their computational cost, even in the presence of
non trivial relationships between the features (e.g. feature B requires feature A to be computed first, the cost of feature C depends on its value).
Such an approach could be particularly important for experiments which aim to study a wide range of processes whose distinguishing features have
very different computational costs, for which a ``by hand'' cascade construction is simply impractical.

\section{Real-time analysis at the LHC: the present}
\label{sec:present}

Having understood how analysis cascades are budgeted, we will now describe the present state of real-time analysis at the LHC. 
We will focus on which features are used to achieve data reduction at each analysis stage, and how they are used. Although
the use of real-time machine learning is still in its infancy at the LHC, by the end of this section you should be able to follow
an LHC real-time analysis cascade and understand where machine learning approaches might improve its performance. 

The dominant data reduction paradigm at the LHC is that the real-time analysis should classify entire raw events as either interesting or 
uninteresting, and keep the former while discarding the latter. The logic behind this approach\footnote{Succintly described in \href{http://www-conf.slac.stanford.edu/ssi/2006/lec\_notes/Sphicas072606.pdf}{http://www-conf.slac.stanford.edu/ssi/2006/lec\_notes/Sphicas072606.pdf} by Paras Sphicas.} is that only a
small fraction of LHC $pp$ collisions produce particles or processes which we are interested in studying, but that when such a particle
or process \textbf{is} produced, we should keep all the available information about the raw event in order to best understand it.
Within this paradigm, the typical process of interest is the production of a particle which is heavy compared to the rest of the raw event: for
example, the Higgs boson which is a focus of ATLAS and CMS analyses has a mass of roughly 125~GeV, while particles containing bottom quarks ($b$-hadrons) which
LHCb was built to study have a mass around 5~GeV. When these particles decay, their mass is converted into momentum transverse to the LHC
beamline, which is the most fundamental quantity used by present LHC real-time analyses. Of course, any data reduction strategy which relies
on large transverse momentum will consequently not be optimal for studying the properties of light particles, and we will return to the implications
of this in Sec.~\ref{sec:future}. 

\begin{figure}
\centering
\includegraphics[width=1.0\linewidth]{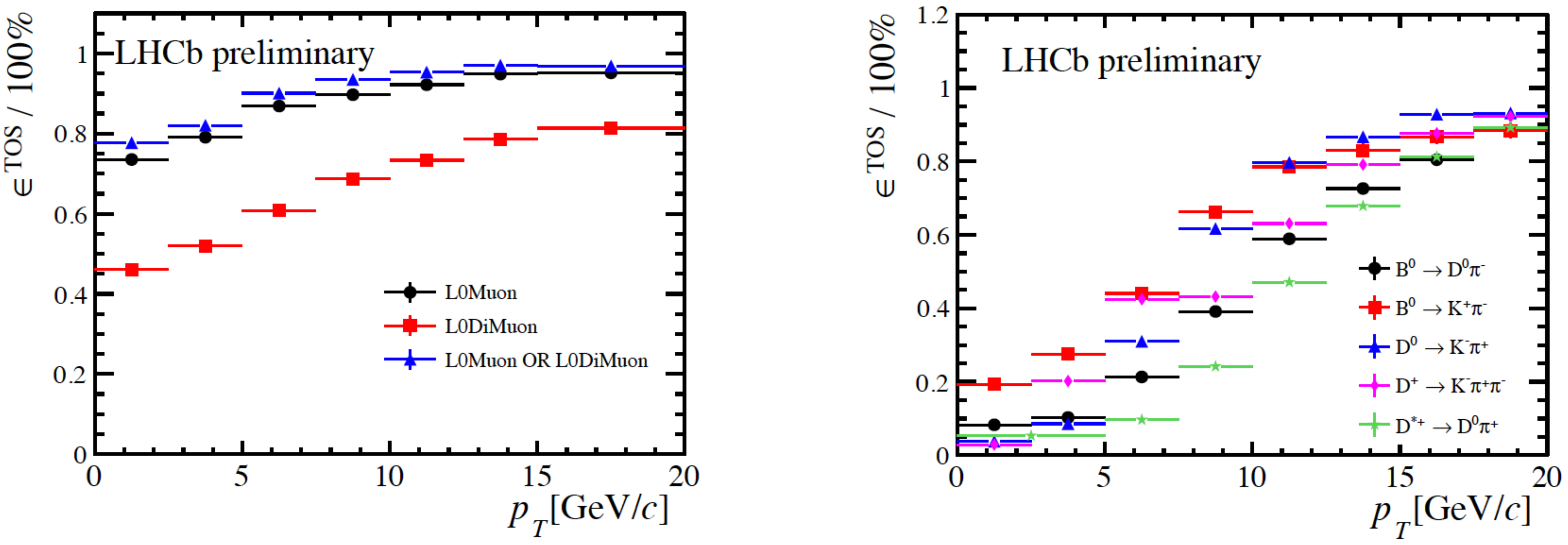}
\caption{The efficiency (true positive rate) of LHCb's muon (left) and hadron calorimeter (right) hardware triggers to select certain processes of interest,
reproduced from~\cite{Albrecht:1644942}.
The efficiency is plotted as a function of the transverse momentum of the decaying particle.
For the muons the process of interest shown is the decay $B\to J/\psi K$, whereas the different processes of interest for the calorimeter are labelled on the plot.}
\label{fig:HardwareTriggerEff}
\end{figure}

The advantage of transverse momentum, or energy, is that they can be quickly calculated using only part of the information in the raw event,
making them ideal quantities for use in hardware triggers\footnote{Transverse energy is a concept used to approximately convert energy measurements
in calorimeters into a measurement of the transverse momentum of the detected particle. Energy is of course a scalar, but if we assume that the particle
is relativistic then its energy is approximately equal to the magnitude of its momentum. Therefore, the absolute measurement of the energy can be converted into
a measurement of transverse momentum by using the vector between the detected calorimeter cluster and the centre of the $pp$ interaction region.}.
In particular, the transverse energy of a photon, electron, or hadron can be computed
from the clusters which they leave in the detector calorimeters, while for muons it is possible\footnote{Muons are special because they are
the only particles which penetrate beyond a detector's calorimeters, and few muons are produced in each $pp$ collision. This means
that the muon chambers have very few hits compared to the rest of the detector, and it is consequently possible to reconstruct muon
trajectories within the constraints of a hardware trigger.}
to reconstruct their trajectories and measure the transverse momentum. Since both clusters and muon segments are highly localized objects,
not only do different subdetectors not have to exchange information, but each subdetector can be split into independently processing regions of interest (ROI),
further parallelizing and speeding up the problem. These two basic quantities form the backbone of all LHC hardware triggers,
with some combinations for specific signals: for example two muons with large transverse momentum signal the presence of a particularly interesting
kind of $b$-hadron decay, while particles of interest to ATLAS and CMS often decay into ``jets'' of high transverse momentum particles grouped
within a narrow cone in solid angle. Each subdetector (hadron calorimeter, electron calorimeter, muon stations, and so on) searches for the specific signature of interest
and signals the presence of any interesting signatures to a global decision unit, which combines the subdetector information and gives the overall
accept or reject decision. The performance of this hardware trigger is shown for the case of LHCb in Fig.~\ref{fig:HardwareTriggerEff}.
While exceptions exist, it is a good general rule of thumb that muon hardware triggers are the most efficient at reducing the data volume,
followed by triggers which look for photon or electron signatures, and finally those triggers which look for high energy hadronic signatures. 

Following this first, hardware, stage, the CMS, ATLAS, and LHCb detectors send all the data about the surviving raw events to farms of CPU servers,
where data from all subdetectors is used to build a complete picture of each raw event. 
All three detectors send around 50~GBs$^{-1}$ of data to the CPU farm for further analysis and reduction.
Once the raw events have been built, the final data reduction is
achieved in the kind of analysis cascade described earlier, and the raw events are sent to permanent storage. I will now describe one kind
of analysis cascade used in the software trigger of LHCb in some detail, which is nevertheless illustrative of the kinds of analysis cascades used
at the LHC generally\footnote{This analysis cascade evolved during the 2010--2012 data taking period, and all performance numbers which I quote should
be taken as illustrative; please refer to~\cite{Albrecht:1644942} for the latest citeable numbers.}.

Our chosen cascade is LHCb's inclusive $b$-hadron trigger, which aims to retain raw-events containing the decays of $b$-hadrons which can
be reconstructed in the LHCb detector. As with our pedagogical example, this cascade is split into a fast analysis (\cite{LHCb-PUB-2011-003})
and a slow analysis\footnote{See \cite{Williams:1323557,Gligorov:1384380}.}; the fast analysis requires around 10--15~ms to reach a decision, and reduces 
the data rate by a factor of 12--20, while the slow analysis requires around 100--200~ms to reach a decision and reduces the data rate by a further factor
of around 40, for a total data rate reduction factor of 480--800. Both the fast and the slow analysis are based around two essential features
of a $b$-hadron decay: the large transverse momentum of the children, and the fact that a $b$-hadron lives longer than many other particles,
and therefore decays at a point in the detector which is well separated from the initial $pp$ collision. 

\begin{figure}
\centering
\includegraphics[width=0.8\linewidth]{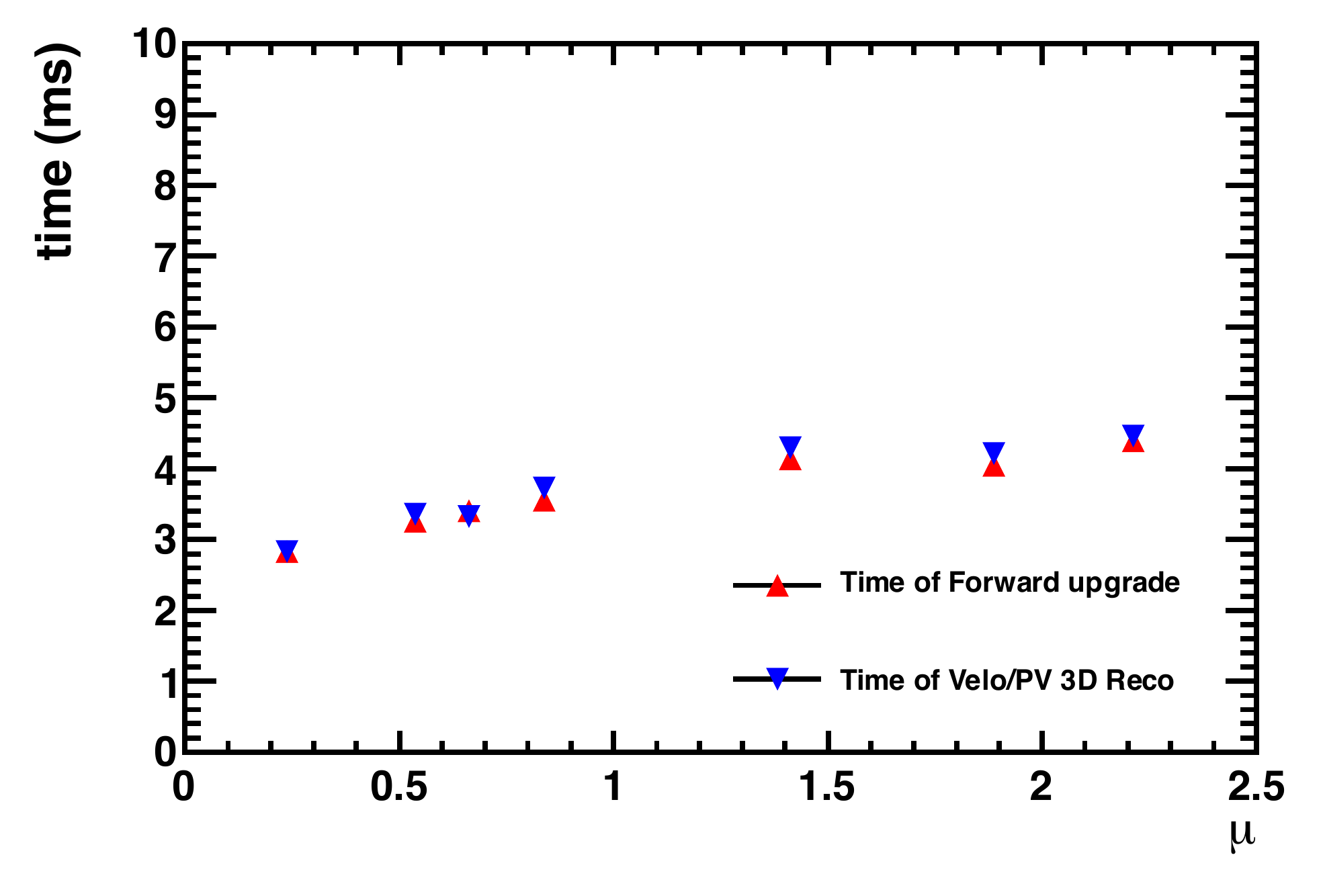}
\caption{The time taken by the VELO and forward reconstructions as a function of the number of $pp$ interactions ($\mu$) in a given
raw LHCb event.}
\label{fig:timings}
\end{figure}

The fast analysis begins by reconstructing all charged particle trajectories in the VELO subdetector. This reconstruction is fast because
there is no magnetic field in the VELO, so that all charged particle trajectories are straight lines, and because the VELO occupancy\footnote{The
occupancy of a subdetector is the fraction of its channels which record a hit from a particle in any given event. In general a smaller
detector occupancy leads to a faster reconstruction because there are fewer ambiguities about which detector hits should be combined into particle
trajectories.} is only around 2$\%$.
This reconstruction already allows those charged particles which are detached from the initial $pp$ collision to be identified,
but this selection alone does not yet allow for entire events to be rejected. For that, we need to identify the second signature of
of $b$-hadron decay: that this detached particle also had a large transverse momentum. We could imagine doing this by matching the VELO
particle trajectories to calorimeter clusters, but this approach is inefficient: the calorimeter has around $20\%$ energy resolution in
the region of interest, and the matching between VELO trajectories and calorimeter clusters also suffers from ambiguities\footnote{The particle
will pass through the magnetic field between the VELO and calorimeter, and since its charge is not known, matching calorimeter clusters must
be searched for on both sides.} and poor resolution. It is much better, but also much more time consuming, to follow the VELO particle trajectory 
through the magnet and match it to hits in the downstream tracker, which gives a $0.5\%$ resolution on the charged particle momentum.
The time to match a VELO track to hits in the downstream tracker is inversely proportional to the track momentum: the higher the momentum, the 
closer the particle trajectory is to a straight line, and the smaller the window in which we have to search for matching hits in the downstream tracker.
The fast analysis therefore does not attempt to find the momentum of all detached VELO tracks, but instead only looks for downstream tracker hits
in a window of transverse momentum $>1.25$~GeV. In this way the transverse momentum which would identify the charged particle as a $b$-hadron decay
product can be used before it has actually been measured. This reduces the timing of this ``forward'' tracking
so it is as fast as the VELO tracking, as shown in Fig.~\ref{fig:timings}; the combination of the detachement and momentum information is sufficient to achieve
the required rate reduction with an efficiency of $>80-90\%$ for most $b$-hadron decays of interest.

Having reduced the data rate in this way, the slow analysis has enough time to perform a more complete event reconstruction. This is also necessary
because a further efficient reduction of the rate can only be accomplished by identifying other charged particles which originated in the $b$-hadron
decay and reconstructing the $b$-hadron decay vertex. The reason for this is that $b$-hadrons are not the only particles which produce
high transvere momentum charged particles detached from the $pp$ collision. In particular, particles containing a charm quark ($c$-hadrons) are
produced 20 times more frequently than $b$-hadrons; while they are around 3 times lighter and have around 3 times shorter lifetimes than $b$-hadrons,
the production rate still makes them a formidable background. In order to efficiently discriminate between the two, the slow analysis
first reconstructs charged particle trajectories in a much looser transverse momentum window of $>0.3$~GeV, in order for the selection
to have access to the majority of $b$-hadron decay products. It then reconstructs 2-,3-, and 4-track vertices, using the track selected
by the fast analysis as a seed; this use of a preceeding selection to seed subsequent ones is common in analysis cascades of this type.

Finally the selection uses a discretized ``Bansai'' Boosted Decision Tree (\cite{Gligorov:2012qt}) to separate the $b$-hadron vertices
from the $c$-hadron vertices and other backgrounds with around $60-70\%$ efficiency. The algorithm was trained
using simulated events for the signal and randomly selected real $pp$ collisions for the background, and the input features
were discretized into between 2 and 20 bins. The width of each bin was chosen to be much larger than the detector resolution for that
specific feature. This serves two purposes: first, since the bins are wider than the detector resolution, events migrate from
one bin to another more rarely even if the detector performance changes; second, the algorithm response can be converted into a one-dimensional
lookup table and hence becomes instantaneous to evaluate. Finally, the features used in the algorithm were explicitly pruned to
a minimum of seven, which maintained $95\%$ of the performance of the complete feature set.
Using this machine learning algorithm allowed a three times greater data reduction for the same
signal efficiency. While this was not strictly the first use of a real-time machine learning algorithm in a high-energy physics
experiment, it is neverthless notable as the first use on such a scale: the majority of papers produced by the LHCb collaboration rely on
this particular algorithm for their data. 

Summarizing the state-of-the art in real-time analysis at the LHC, we see the following key features: the goal of data reduction is to select
interesting whole raw events and discard others; all experiments rely on hardware triggers
with a fixed latency for the first part of their data reduction; and the final data reduction is performed using analysis cascades
on farms of CPU servers. We shall now see how these features will evolve, along with the LHC itself, over the next 10--20 years.

\section{Real-time analysis at the LHC: the future}
\label{sec:future}

The key problem facing real-time analyses at the LHC over the next decades will be the ever increasing data rate which they will have
to reduce. In summarizing the current thinking on this topic, I will draw heavily on material presented at the 2nd ECFA High-Luminosity LHC workshop
in Aix-Les-Bains\footnote{The material is available at
\href{https://indico.cern.ch/event/315626/contributions}{https://indico.cern.ch/event/315626/contributions}}.
In short, the ALICE input data rate will rise to 1~TBs$^{-1}$ in 2019--2020, while the LHCb input data rate will rise to 5~TBs$^{-1}$ at the same time.
ATLAS and CMS will increase their input rates later, around 2024--2027, reaching around 150~TBs$^{-1}$. In all cases, the increase will mean a greater number
of $pp$ interactions per event, and a consequently greater event complexity and slower reconstruction times. In the case of
ATLAS and CMS in particular, the overall volume of data which will have to be processed every year is 600~EB, on a par with the biggest commercial
data processing tasks. On the other hand, the projected computing budgets of the collaborations will not increase over this period,
so that the same or even harsher data reduction targets will have to be met in this more complex environment. 

Before describing the strategies of the individual collaborations, it should be said that they all rely on a continuation of Moore's law in its
current, parallel, incarnation. Intel's current chip roadmap gives us good reason to think that the tick-tock model of improved microarchitecture
and die shrinking will allow Moore's law to continue to hold well into the 2020s, and all collaborations assume that the currently observed
gains in performance per unit cost of 25--35$\%$ per annum can be maintained over the next decade. Nevertheless, even allowing for such improvements,
improvements of between a factor 5 and 10 will have to be found in algorithm speed, by a combination of smarter algorithms and
by making better use of parallel architectures.

The most radical approach to future data reduction is that of ALICE, who begin by rejecting the premise that data reduction should be achieved
by keeping interesting whole events. As ALICE collides lead ions, not protons, all collisions are interesting to some of their analyses; however,
not all particle trajectories within a given collision are equally interesting. For this reason ALICE aim to achieve the required factor 100 data
reduction by performing a lossy data compression within events. This data compression is achieved by an interleaved cascade of reconstruction, data compression,
detector calibration, further reconstruction, compression, etc. Furthermore, ALICE will use a highly heterogeneous ``Online-Offline'' data processing,
bringing together data reduction and compression in the readout electronics of subdetectors, CPU and GPU farms, as well as in GRID/Cloud facilities.

The key the ALICE approach is the ability to transfer the entire 1~TBs$^{-1}$ from the detector, which will easily be possible in 2019-2020. It will
also be possible in the case of LHCb's 5~TBs$^{-1}$, and LHCb will therefore also move to a real-time analysis architecture without any hardware trigger. 
Unlike ALICE, LHCb does not intend to perform data reduction purely through event compression: some events will still be labelled as uninteresting
and rejected, some events will be kept compressed in line with the ALICE model, and some events will be kept whole. 
Nevertheless, the fraction of events which can be labelled as uninteresting will be significantly smaller~\cite{Fitzpatrick:1670985} than in the case
of the current LHCb experiment. This is because LHCb is able to reconstruct very low mass, and low momentum, particles, which are currently rejected
by its momentum-based hardware trigger. With the hardware trigger removed, however, it will in principle be possible to analyse all these lighter signals,
assuming that efficient analysis cascades which fit into the real-time computing resources can be designed. It will be especially important to use
the RICH detectors in the real-time analysis cascades, in order to distinguish pions, kaons, and protons from each other, because a given
particle's decay can be more or less interesting to study depending on the precise mixture of the particle types which it decayed into.
For this reason, LHCb plans to make extensive
use of the two stage buffering model described earlier, and perform a continuous real-time detector alignment and calibration in between these two stages;
particularly important for the gas-based RICH detectors whose performance is highly sensitive to the temperature and pressure in the cavern and varies
on an hourly basis.
The possibility of having further buffering stages in the GRID/cloud, as ALICE, is also under consideration. The hardware which will be used for event
processing is not yet defined, and both CPU, GPU, and XeonPhi-like architectures are under review. The lack of a hardware trigger, however, means that LHCb
will have maximum flexibility to wait and see how the processing hardware evolves before committing to a specific architecture quite close to 2019--2020.

In the case of CMS and ATLAS, it is not possible to imagine transfering 150~TBs$^{-1}$ from the detectors even in the mid-2020s. The key problem is the
detector layout, which for CMS and ATLAS is a sort of onion hermetically covering the full $4\pi$ of solid angle around the collision point. In this
case the readout cables for the subdetectors have to run through the detector volume, which means that their mass must be kept to a minumum in order
not to scatter the particles which we are trying to measure and spoil the detector resolution. Of course the network infrastructure to transfer 150~TBs$^{-1}$
would also be challenging, but probably not entirely impossible if the cabling constraint didn't exist. For this reason both ATLAS and CMS will maintain
hardware triggers, which will reduce the data rate by a factor of between 40 and 75, not coincidentally quite similar to the ALICE and LHCb data rates.
In addition, both ATLAS and CMS will largely maintain the paradigm of reducing their data volume by selecting interesting events whole,
although mixed strategies in which certain kinds of events are kept in a compressed format are also under consideration.
These ATLAS and CMS hardware triggers will be augmented by specialized electronics allowing the reconstruction of non-muon charged particle
trajectories, a first at anything like these data rates in high energy physics. These so-called track-triggers will allow for a more efficient data reduction,
before the data is sent to processing farms similar to the ones used by ALICE and LHCb in order to finish the job. As with LHCb, the precise hardware
which will be used in these processing farms is still under investigation.

Before concluding, it is worth remarking on an important trend which will likely become more prominent in the coming years at the LHC. This
is the use of real-time multivariate analysis, as already pioneered by LHCb. While high-energy physics has traditionally been slow to adapt the most
advanced data science analysis paradigms, the majority of papers being published by the LHC collaborations use multivariate methods (whether decision trees,
neural networks, support vector machines, deep learning...) in their final analysis, and these methods are also migrating more and more into the real-time world.
In particular, such methods will likely become more important in the case of ALICE and LHCb, where a majority if not all of the events are interesting
for some analysis, and the goal of real-time analysis is not to distinguish between interesting signals (particles of interest) and uninteresting backgrounds (other particles),
but to classify different kinds of interesting signals.

\section{Conclusions}

These proceedings have hopefully given you an insight into the current state of real-time data analysis at the LHC, and much of what has been
presented is applicable to any physics experiment whose data rate is too high to allow all the data to be saved to long term storage. The general
trend is towards more and more heterogenous data processing and analysis methods, both in terms of the algorithms used to analyse the data and
the computer hardware they run on. The LHC experiments are already among the biggest data processing facilities in the world, and will continue
to evolve as such over the next 10--20 years; the advances made in real-time analysis at the LHC will therefore necessarily follow, and hopefully feed back into,
similar advances in commercial applications and other scientific domains.


\acks{Mike Sokoloff wrote the script used to illustrate deferred triggering.
I am very grateful to Marco Cattaneo, Graeme Stewart, and Mika Vesterinen, as well as three anonymous reviewers,
for reading and helping to improve this manuscript.}

\bibliography{my_bib}

\end{document}

%% file: lhcb-symbols-def.tex
\def\lhcb {\mbox{LHCb}\xspace}

\def\lhc    {\mbox{LHC}\xspace}

\ifthenelse{\boolean{uprightparticles}}%
{

 \def\PDelta      {\ensuremath{\Delta}\xspace}                 
 \def\PXi      {\ensuremath{\Xi}\xspace}                 
 \def\PLambda      {\ensuremath{\Lambda}\xspace}                 
 \def\PSigma      {\ensuremath{\Sigma}\xspace}                 
 \def\POmega      {\ensuremath{\Omega}\xspace}                 
 \def\PUpsilon      {\ensuremath{\Upsilon}\xspace}                 
 
 \def\PB      {\ensuremath{\mathrm{B}}\xspace}                 
                  
 \def\PD      {\ensuremath{\mathrm{D}}\xspace}

 \def\PK      {\ensuremath{\mathrm{K}}\xspace}

 \def\Pi      {\ensuremath{\mathrm{i}}\xspace}

}
{

 \mathchardef\PDelta="7101
 \mathchardef\PXi="7104
 \mathchardef\PLambda="7103
 \mathchardef\PSigma="7106
 \mathchardef\POmega="710A
 \mathchardef\PUpsilon="7107
                  
 \def\PB      {\ensuremath{B}\xspace}                 
                  
 \def\PD      {\ensuremath{D}\xspace}

 \def\PK      {\ensuremath{K}\xspace}

 \def\Pi      {\ensuremath{i}\xspace}

}

  \def\Kbar  {\kern 0.2em\overline{\kern -0.2em \PK}{}\xspace}

  \def\Dbar    {\kern 0.2em\overline{\kern -0.2em \PD}{}\xspace}

\def\Bbar    {\ensuremath{\kern 0.18em\overline{\kern -0.18em \PB}{}}\xspace}

  \def\Y#1S{\ensuremath{\PUpsilon{(#1S)}}\xspace}

\def\Lbar {\ensuremath{\kern 0.1em\overline{\kern -0.1em\PLambda}}\xspace}

\def\to                 {\ensuremath{\rightarrow}\xspace}

\def\AT#1     {\ensuremath{A_{\mathrm{T}}^{#1}}\xspace}           

\def\C#1      {\ensuremath{\mathcal{C}_{#1}}\xspace}                       
\def\Cp#1     {\ensuremath{\mathcal{C}_{#1}^{'}}\xspace}                    
\def\Ceff#1   {\ensuremath{\mathcal{C}_{#1}^{\mathrm{(eff)}}}\xspace}        
\def\Cpeff#1  {\ensuremath{\mathcal{C}_{#1}^{'\mathrm{(eff)}}}\xspace}       
\def\Ope#1    {\ensuremath{\mathcal{O}_{#1}}\xspace}                       
\def\Opep#1   {\ensuremath{\mathcal{O}_{#1}^{'}}\xspace}                    

\newcommand{\tev}{\ifthenelse{\boolean{inbibliography}}{\ensuremath{~T\kern -0.05em eV}\xspace}{\ensuremath{\mathrm{\,Te\kern -0.1em V}}\xspace}}
\newcommand{\gev}{\ensuremath{\mathrm{\,Ge\kern -0.1em V}}\xspace}
\newcommand{\mev}{\ensuremath{\mathrm{\,Me\kern -0.1em V}}\xspace}
\newcommand{\kev}{\ensuremath{\mathrm{\,ke\kern -0.1em V}}\xspace}
\newcommand{\ev}{\ensuremath{\mathrm{\,e\kern -0.1em V}}\xspace}
\newcommand{\gevc}{\ensuremath{{\mathrm{\,Ge\kern -0.1em V\!/}c}}\xspace}
\newcommand{\mevc}{\ensuremath{{\mathrm{\,Me\kern -0.1em V\!/}c}}\xspace}
\newcommand{\gevcc}{\ensuremath{{\mathrm{\,Ge\kern -0.1em V\!/}c^2}}\xspace}
\newcommand{\gevgevcccc}{\ensuremath{{\mathrm{\,Ge\kern -0.1em V^2\!/}c^4}}\xspace}
\newcommand{\mevcc}{\ensuremath{{\mathrm{\,Me\kern -0.1em V\!/}c^2}}\xspace}

\def\mum  {\ensuremath{{\,\upmu\rm m}}\xspace}

\def\gsim{{~\raise.15em\hbox{$>$}\kern-.85em
          \lower.35em\hbox{$\sim$}~}\xspace}
\def\lsim{{~\raise.15em\hbox{$<$}\kern-.85em
          \lower.35em\hbox{$\sim$}~}\xspace}

\def\tell1  {TELL1\xspace}
\def\ukl1   {UKL1\xspace}

